\documentclass[10pt]{sig-alternate-05-2015}
	
\usepackage[
left=1.9cm,top=1.9cm,right=1.9cm, bottom=2.55cm
]{geometry}

\usepackage[table]{xcolor}
\usepackage{url}
\usepackage{wireless-utils}
\usepackage{epstopdf}
\usepackage{float}
\usepackage[pdf]{pstricks}
\usepackage{amsfonts,amssymb}
\usepackage{amsmath,bm}
\usepackage{graphicx}
\usepackage{array}
\usepackage[off]{auto-pst-pdf}
\usepackage[small]{subfigure}
\usepackage{balance}
\usepackage{times}
\usepackage{paralist}
\usepackage{booktabs}
\usepackage{algorithm}
\usepackage{algorithmic}
\usepackage[off]{auto-pst-pdf}
\usepackage[small]{subfigure}

\usepackage{etoolbox}
\newcommand{\tabincell}[2]{\begin{tabular}{@{}#1@{}}#2\end{tabular}}
\usepackage{enumitem}
\setitemize{noitemsep,topsep=2pt,parsep=2pt,partopsep=2pt}

\begin{document}

\CopyrightYear{2016} 
\setcopyright{acmcopyright}
\conferenceinfo{CoNEXT '16,}{December 12-15, 2016, Irvine, CA, USA}
\isbn{978-1-4503-4292-6/16/12}\acmPrice{\$15.00}
\doi{http://dx.doi.org/10.1145/2999572.2999584}

\title{Passive Communication with Ambient Light}

\numberofauthors{3}
\author{
	\alignauthor 
	Qing Wang\titlenote{{Part of the work was performed when the author was PhD student at IMDEA Networks Institute, Madrid, Spain.}}\\
	\affaddr{TU Delft}\\
	\affaddr{Delft, the Netherlands}\\
	\affaddr{{q.wang-5\\@tudelft.nl}}
	\alignauthor
	Marco Zuniga \\
	\affaddr{TU Delft}\\
	\affaddr{Delft, the Netherlands}\\
	\affaddr{m.a.zunigazamalloa\\@tudelft.nl}
	\alignauthor
	Domenico Giustiniano \\
	\affaddr{IMDEA Networks Institute}\\
	\affaddr{Madrid, Spain}\\
	\affaddr {domenico.giustiniano\\@imdea.org}
}
\maketitle

\abstract
In this work, we propose a new communication system for illuminated areas, indoors and outdoors.  
Light sources in our environments --such as light bulbs or even the sun-- are our signal emitters, but we do not modulate data at the light source. We instead propose that the environment itself modulates the {ambient} light signals: if mobile elements `wear' patterns consisting of distinctive reflecting surfaces, single photodiode could decode the disturbed light signals to read \emph{passive} information. Achieving this vision requires a deep understanding of a new type of communication channel. Many parameters can affect the performance of passive communication based on visible light: the size of reflective surfaces, the surrounding light intensity, the speed of mobile objects, the field-of-view of the receiver, to name a few. In this paper, we present our vision for a passive communication channel with visible light, the design challenges and the evaluation of an outdoor application where our receiver decodes information from a car moving at 18 km/h.


\section{Introduction} \label{sec_intro}
Between the late 1960s and the late 1970s, trains and other rolling stock in North America were monitored based on the standard KarTrak~\cite{nelson1997punched}. This standard consisted of `big' barcodes ($\approx$15$\times$45\,cm) attached to wagons and readers the size of a shoebox that had a ``light source and photo-multipliers for sensing the reflected light from the barcode''\footnote{KarTrak is the predecessor of the small and highly effective barcode systems that are now pervasive in supermarkets.}.

\begin{figure}[t!]
	\centering
	\includegraphics[width=0.65\columnwidth]{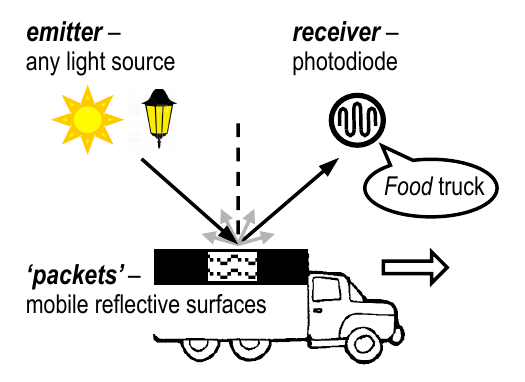}
	\vspace{-3mm}
	\caption {An application of our passive channel.}
	\label{fig_basic_channel}
	\vspace{-4mm}
\end{figure}

In this paper, we propose to look back at the idea of KarTrak, and barcodes in general, but from a different perspective. Instead of building sophisticated lights into the reader, we exploit existing light sources such as the sun. The reader is then reduced to an inexpensive and energy-efficient `tiny box' consisting of simple off-the-shelf photodiodes.
By deploying a large number of these tiny boxes, we could gain valuable information about our environment. For example, vehicles can use distinctive reflective surfaces to encode information such as the type of cargo or id, as shown in Fig.~\ref{fig_basic_channel}. Emergency, treatment, and housekeeping trolleys could embed codes to inform their physical locations in a hospital. {Furthermore, objects' intrinsic reflective surfaces, e.g., car's roof/windshield, the shape of people, etc., can be exploited by these tiny boxes for applications such as localization.}

A key distinctive feature of our proposal is \emph{sustainability}: it has a low footprint across many dimensions. First, \emph{infrastructure}. A large part of our system builds on top of existing infrastructure: there is no need for extra light sources or electronic transmitters on objects. Second, \emph{energy}. Cameras can also provide a passive monitoring infrastructure (by reading QR codes) {and have been used for human-computer interaction}~\cite{Wachs:2011}, visual MIMO~\cite{Ashok:2010} and localization~\cite{Kuo2014}. But cameras consume orders of magnitude more energy than simpler photodiodes: upwards of 1000\,mW~\cite{Chen2013cameraenergy} vs 1.5\,mW (power consumption of the photodiode used in our system{{\footnote{{The photodiode is TI OPT101 (\url{https://goo.gl/TDsXS3}). Its power consumption is measured in our lab.}}}}).
This low power requirement would enable a small solar panel --the size of a credit card-- to harvest enough energy from the surrounding lights for our system to work autonomously. Third, \emph{cost}. {Cameras pose not only threats to user's privacy but} can be far more expensive (depending on specific features) than single photodiodes. {In order to move into higher efficiency low carbon markets, modern ICT technology should give a centric role to sustainability in the design space}~\cite{SMART2020}.

The performance of our system however depends on many parameters that are outside the control of the system itself. Unlike traditional barcode readers, which adopt coherent light (laser), we have no influence over the location, characteristics or illuminance power of surrounding lights. We have no influence either over the speed of passing objects: \textit{we can not slow down an object to read its code}. All these dynamics affect the system's ability to decode data. To gain a deeper insight on our system's performance, {we propose} \emph{a new passive communication channel.} In the next section, we describe the main components of our communication channel and highlight its novelty compared to related work.

\section{New Communication Channel}

The basic functional blocks of a digital communication system --the information source, the transmitter, the channel, and the receiver-- are the foundation of today's wireless communication~\cite{proakis2007}. The idea behind this well-established concept is that information sources should use the transmitter to modulate data, so it can travel through the channel and be decoded by the intended receiver.

\begin{figure}[t!]
	\subfigure[Flowchart of the classic communication (top) and our passive communication system (bottom).]
	{\includegraphics[width=0.66\columnwidth]{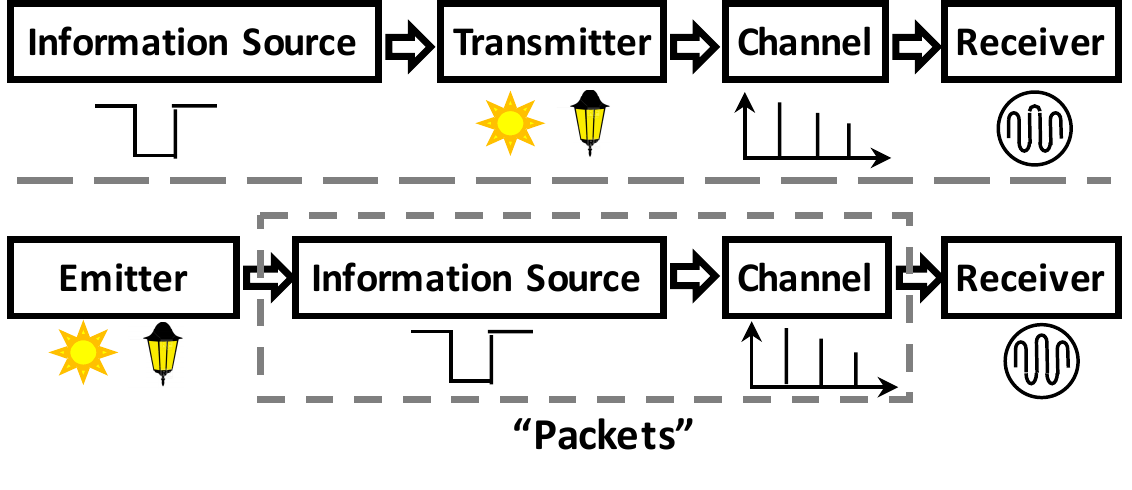}\label{fig_classic_new}}
	\hfill
	\subfigure[FoV effect.]
	{\includegraphics[width=0.33\columnwidth]{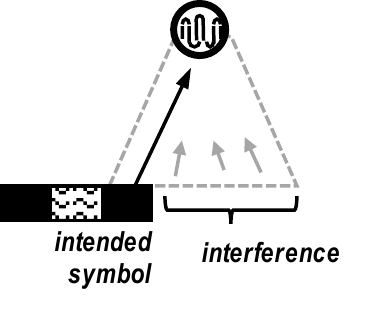}\label{fig_FOV}}
	\vspace{-5mm}
	\caption{Communication system (a) and FoV effect (b)}
	\vspace{-3mm}
\end{figure}

We propose a new flowchart for a \emph{passive} communication system with low carbon footprint of electronics based on visible light, cf. Fig.~\ref{fig_classic_new}. {Our work differs from recent sensing systems {leveraging visible light communication}\footnote{In Visible Light Communication (VLC) light sources are modulated at a high speed to transmit information without affecting the illumination perceived by the user.} for human motions}~\cite{Li:2015-mobicom,Li:2016-mobisys}, mobile interaction~\cite{Zhang:2015}, and localization~\cite{Li2014NSDI,Kuo2014} on the basic fact that \emph{we do not use modulated light sources}. Instead, we embed data into reflective objects for passive communication.

The system is composed of three basic block elements:
\begin{itemize}
	\item \emph{Emitters}, which could be any \emph{simple unmodulated} light source such as standard luminaries or the sun.
	\item \emph{Receivers}, which are tiny boxes containing at least one photodiode to measure the impinging light intensity.
	\item \emph{Surfaces}, which reside at mobile objects and contain different \emph{reflective materials}.
\end{itemize}

We \textit{refer to these surfaces as `packets'}. 
The reflective materials used to encode information could be anything, from a pure mirror (strong reflection, {low power loss}, perpendicular to the incident light), to a dark and rugged cloth (minimal reflection, {high power loss}, scattered in all directions). As the object moves, the intensity of the reflected light rays changes, altering the amount of light impinging towards a particular receiver. This disturbed reflected light can then be detected and decoded by the receiver. Note that the power loss of this communication channel is a function of the reflection coefficient of the reflective material. This simple system could \emph{enable the monitoring of events and activities in a passive manner}. 

{An important aspect of our system is that light are modulated by reflective surfaces. This concept is inspired by backscatter communication, where passive tags modulate the electromagnetic waves produced by external sources. This technique has been traditionally used by RFID tags}~\cite{want2004enabling}, and recently applied to other radio technologies, such as Wi-Fi~\cite{Kellogg:2014} and TV signals~\cite{Liu:2013}. In the same way radio-backscatter exploits the surrounding radio waves, we want to exploit visible light. However, our tags (packets) are not electronic devices and our system has to deal with visible light waves, which have completely different properties compared to radio waves. Wireless barcode~\cite{Moshir2014} has been proposed to embed information into infrastructure such as building walls. The barcode was built with material (e.g., copper/cement) into a shape of square-wave. Data was modulated through the barcode's reflection of electromagnetic waves. This barcode did not have electronic devices, similar to our tag. However, \cite{Moshir2014} required expensive dedicated radio signals impinging on it (the unit cost\footnote{{See \url{https://goo.gl/jQacLJ}}} is $\$220,000$), while ambient light is sufficient for our tag to work (our prototype costs around 50 dollars). Only recently, Retro-VLC explored backscattering with visible light communication~\cite{Li:2015}. Retro-VLC created a bidirectional link between an \emph{active} light source and a \emph{static} semi-passive with LCD shutters. The LCD shutters were used to change the reflected signal in time. Our work differs from Retro-VLC in two important ways, we explore backscattering for \emph{unmodulated} light sources and \emph{mobile} nodes. \emph{LCD shutters however can complement our work}, at an increased carbon footprint, by enabling mobile objects to change the reflected information (code) in time.

\section{Challenges}\label{sec:challenges}

Since we have no control over the location or illuminance power of light sources,  the receiver's ability to decode information depends strongly on its field-of-view (FoV) and distance to the mobile object. A wide FoV provides a wider coverage but it also exposes the receiver to more interference, as shown in Fig.~\ref{fig_FOV}. A narrow FoV provides the opposite trade-off: a higher signal-to-interference ratio at the expense of having a limited coverage. Regarding the distance to the mobile object, increasing this distance is detrimental for two reasons: (i) the signal strength of visible light waves decrease exponentially with distance (like radio waves), and (ii) longer distances increase the area covered by the FoV, which as stated before can add interference. With this basic understanding of the communication channel, we now describe the most relevant design challenges.

\textbf{Channel capacity.} Communication systems need to quantify the data rate of their channels. For us, this means understanding how different parameters --such as the FoV, distance to mobile objects, width of reflective surfaces and the speed of mobile objects-- affect the capacity of the passive channel.

\textbf{Channel distortions.} Similar to radio systems, our channel will be exposed to distortions. For example: fog, humidity, dirt on top of the reflective surfaces and variable speeds of the mobile object will be commonplace phenomena affecting the incoming signal and making it harder to decode. 

{\textbf{`Packet' collisions.} Until now we assume that a single object moves under the FoV of a receiver. In some applications this will not hold. If several objects move under the same FoV, the incoming signal will be the sum of multiple `overlapping' symbols. Leading in fact to the equivalent of packet collisions in traditional radio communication systems.}

\textbf{Noise floor.} Communications based on visible light need to cope with the fact that surrounding light intensity can change significantly. These `noise floor' changes can easily saturate a photodiode, which make links disappear abruptly.

In Section~\ref{sec_evalu}, we look at each of these challenges from an empirical viewpoint. We showcase the problem and present our solutions. In Section~\ref{sec_app}, we evaluate our passive system through an outdoor vehicle application.

\section{Decoding Information: \\ An Empirical Approach} \label{sec_evalu}
We empirically investigate the challenges described in Section~\ref{sec:challenges}. In our evaluation, we use the following components:

{\bf{Emitter}:} we use three different types of light sources: an LED lamp, ceiling fluorescent lights, and the sun. In this section, most of our results are based on the LED lamp because it allows us to have controlled and repeatable scenarios.  In Section~\ref{sec_app}, we use the sun as light source. 

\begin{figure}[t!]
	\begin{minipage}{0.36\columnwidth}
		\centering
		\includegraphics[width=\columnwidth]{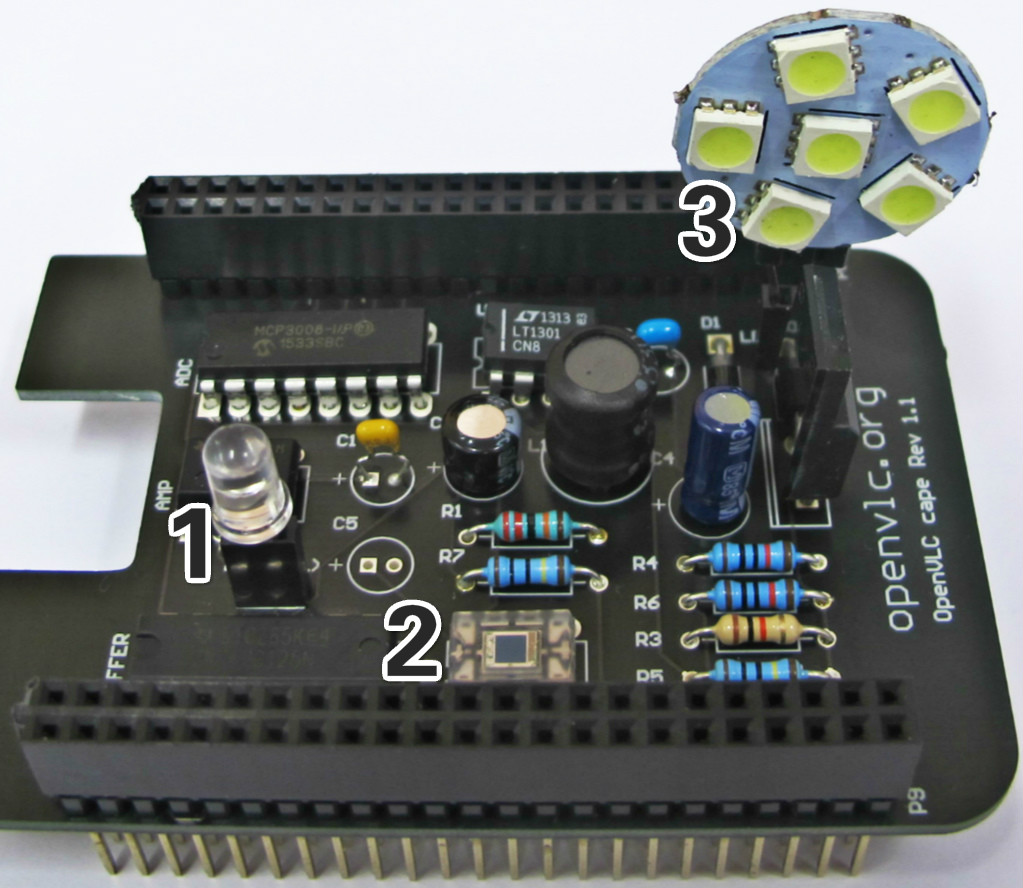}
	\end{minipage}
	\hfil
	\begin{minipage}{0.6\columnwidth}
		\resizebox{\columnwidth}{!}{
			\begin{tabular}{|l|l|}
				\hline
				\textbf{Model} & \textbf{Description} \\  \hline
				\tabincell{l}{HLMP-EG08\\-YZ000} & \tabincell{l}{5~mm red LED (used as a \\receiver)}     
				\tabularnewline  \hline
				OPT101 & {photodiode}         
				\tabularnewline  \hline
				74HCT244N & {tri-state buffer}
				\tabularnewline  \hline
				LM358N & amplifier   
				\tabularnewline  \hline
				MCP3008 & \tabincell{c}{analog-to-digital converter}
				\tabularnewline  \hline
				ADG444 & multiplexer
				\tabularnewline  \hline
		\end{tabular} }
	\end{minipage}
	\caption{The evaluation board runs a Debian Linux and the open-source OpenVLC driver. We receive data using either the low-power LED (1) or the photodiode (PD) (2). Key used electronic devices are listed on the right.}
	\vspace{-3mm}
	\label{fig:openvlc}
\end{figure}

{\bf{Receiver}}: we use and customize the OpenVLC platform~\cite{Wang2015wcm}. The board is depicted in Fig.~\ref{fig:openvlc}, highlighting the two optical receivers. Given that our channel relies on passive sources, we \emph{only} use the OpenVLC interfaces related to reception (i.e., the optical components (1) and (2) in Fig.~\ref{fig:openvlc}). 

Before proceeding with our insights, we first introduce some basic information of our system: the encoding of data and the packet format.

{\bf Coding}. Our encoding process is passive and is performed independently by each moving object. 
In the experiments, we use the following materials to encode information:
\begin{itemize}
	\item Aluminum tape, which has a relatively high reflection coefficient and low diffused reflections (to represent the symbol \texttt{HIGH});
	\item Black paper napkins, which have a lower reflection coefficient and higher diffused reflections (to represent the symbol \texttt{LOW}).
\end{itemize}
The {\it symbol width}, defined as the width of the material representing a symbol, remains constant within a packet, but different packets can have different symbol widths.
To enable an easy and stable decoding at the receiver, we use Manchester codes: a `0'-bit is mapped to \texttt{HIGH-LOW}, and a `1'-bit is mapped to \texttt{LOW-HIGH}.

\begin{figure}[t]
	\centering
	\includegraphics[width=.9\columnwidth]{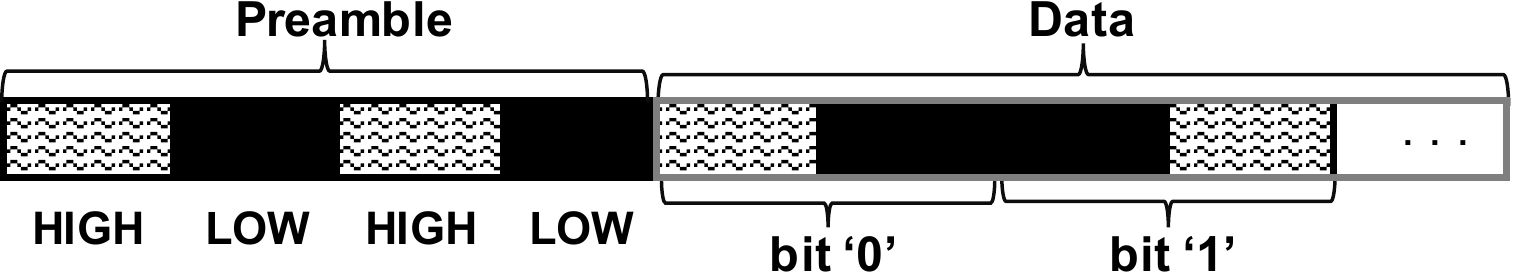}
	\vspace{-2mm}
	\caption {Packet format: Preamble + Data.}
	\label{fig_pkt_format}
	\vspace{-4mm}
\end{figure}

{\bf Packet format.}   
Each packet has two fields: preamble and data, as shown in Fig.~\ref{fig_pkt_format}. The preamble is fixed and consists of four symbols \texttt{HIGH-LOW-HIGH-LOW}. The detection of the preamble requires no \emph{a-priori} calibration, as it will be described later. The Data field comes after the preamble and includes $2N$ symbols, representing the modulated $N$-bit data.

\subsection{Channel Capacity: The Ideal Scenario} \label{sec_ideal_scenarios}

In this subsection, we quantify the channel's capacity. All of our experiments follow the basic setup illustrated in Fig.~\ref{fig_basic_channel}, a photodiode above a passing mobile object. First we describe our decoding method. Then, we assess the impact that the symbol width and the receiver's height have on the channel's throughput.

\textbf{Decoding.} The receiver decodes information based on the Received Signal Strength (RSS). The RSS is perceived as a sequence of  \texttt{HIGH} and \texttt{LOW} symbols. In this subsection we assume that no distortions are present (no dirt or fog) and that objects move at constant speed while passing under the FoV of the receiver. In the next subsections we propose different mechanisms to overcome signal distortion and interference.

Unless stated otherwise, we use an LED lamp as light source and the experiments are carried in an office where no other sources of light are present: the blinds are closed and light bulbs are turned off. These constraints are removed in our outdoor experiments. The ground plane where objects move is covered with black papers, to resemble tarmac. To obtain binary data out of the RSS signal, we use two thresholds: one for the magnitude of the RSS signal $\tau_r$ (to distinguish whether a symbol is \texttt{HIGH} or \texttt{LOW}) and one for the time length $\tau_t$ (to estimate the duration of a symbol). The thresholds are obtained on a per-packet basis and do not require calibration. These thresholds need to be \emph{highly adaptive} because we do not modulate information with a common transmitter, but we rather let each packet determine its own parameters: symbol width, materials used and speed.

\begin{figure}[t]
	\subfigure[Data=`00' (`HLHL')] {\includegraphics[width=0.49\columnwidth]{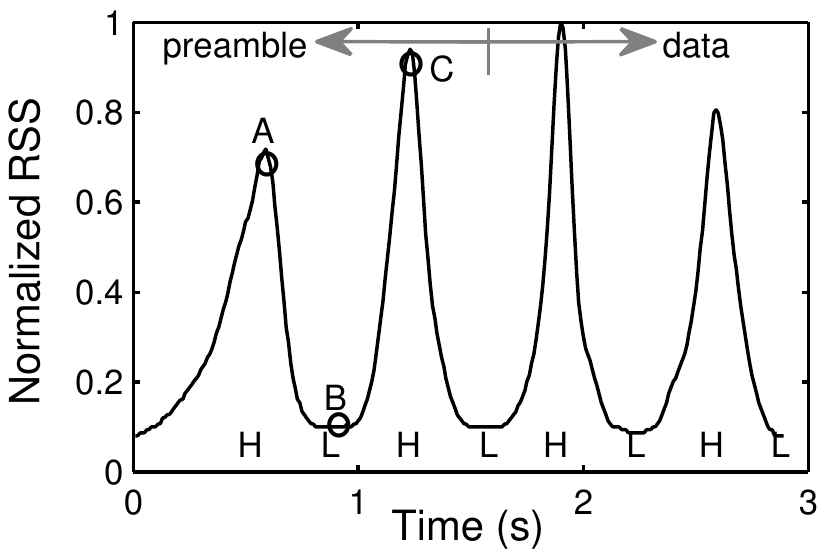}}
	\hfill%
	\subfigure[Data=`10' (`LHHL')] {\includegraphics[width=0.49\columnwidth]{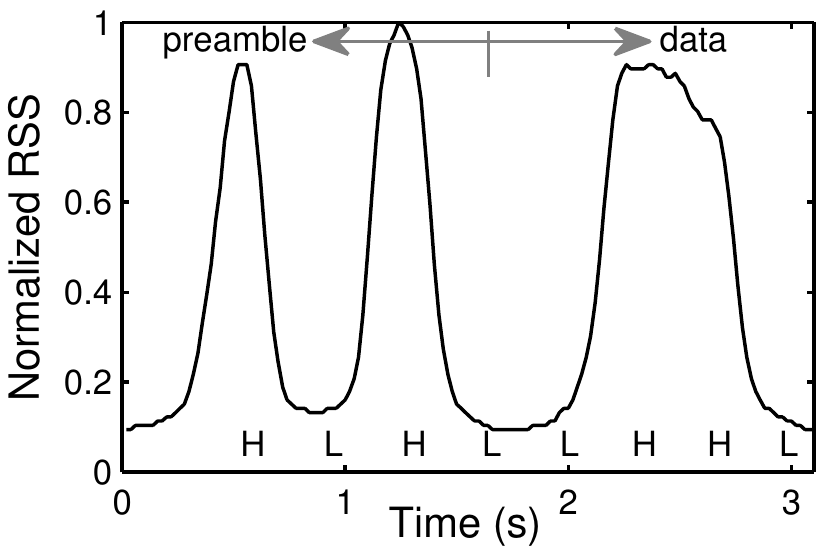}}
	\vspace{-2mm}
	\caption {Received signals in an ideal scenario.} 
	\vspace{-3mm}
	\label{fig_two_ids}
\end{figure}

We first detect the first two peaks and the first valley present in the preamble, points A, B and C in Fig.~\ref{fig_two_ids}(a). Denoting the tuple $<r_i,t_i>$ as the RSS value and timestamp for point $i$, the thresholds for the magnitude and period are defined as
\begin{align*}
	\tau_r = \frac{(r_A - r_B) + (r_C - r_B)}{2} ; \
	\hfill
	\tau_t = \frac{(t_B - t_A) + (t_C - t_B)}{2}.
\end{align*}

With these thresholds, subsequent RSS measurements are grouped together in windows of length $\tau_t$. If the maximum value in a window is above $\tau_r$, we declare the symbol to be \texttt{HIGH}; otherwise, we declare it to be \texttt{LOW}. Figure~\ref{fig_two_ids} depicts the RSS for two packets carrying a two-bit (four-symbol) payload: ``00'' (left) and ``10'' (right). Both packets have the same symbol width (3~cm). The emitter and receiver are at a height of 20~cm from the workplane, and the distance between emitter and receiver is 12~cm. For this scenario, the received signals are clear. It is therefore easy for our decoding method to get accurate thresholds values (based on the preamble), and consequently, to decode information.

\begin{figure}[!t]
	\subfigure[Height vs. symbol width]{\includegraphics[width=0.49\columnwidth]{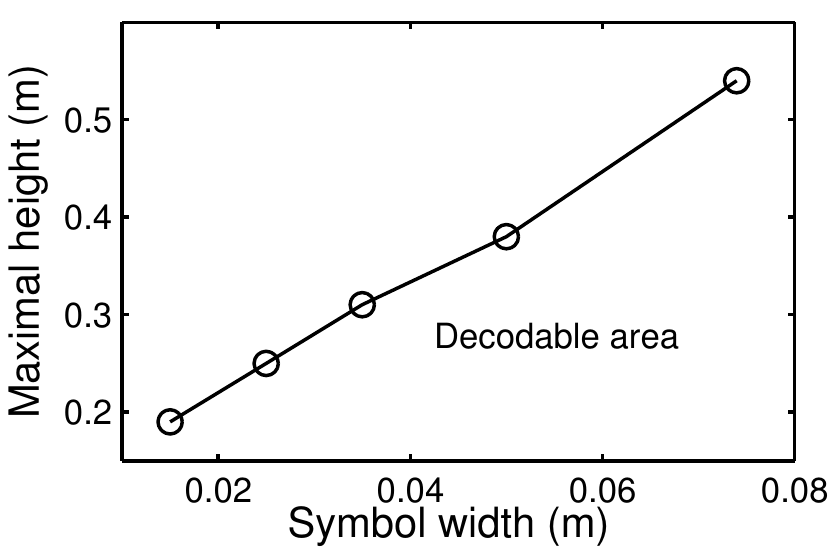}}
	\hfill%
	\subfigure[Height vs. throughput] {\includegraphics[width=0.49\columnwidth]{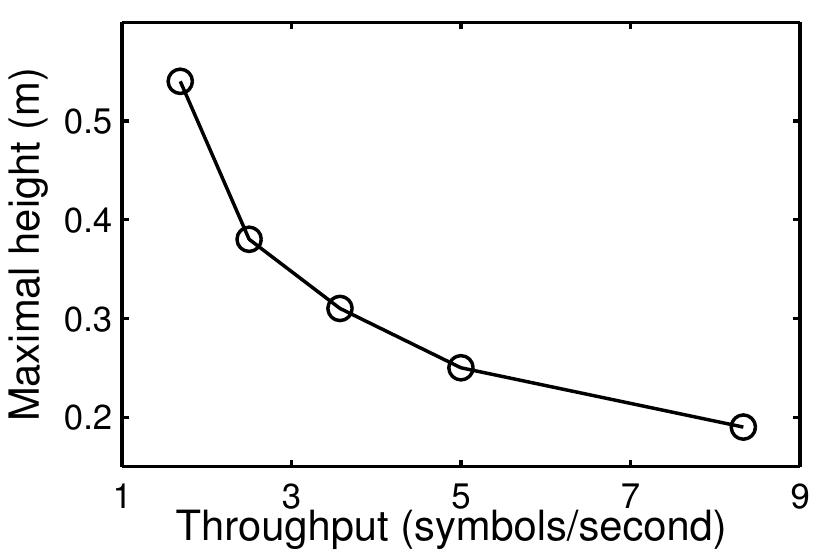}}
	\vspace{-2mm}
	\caption {Maximal height vs. symbol width and system \newline throughput. Packet's moving speed is 8cm/s.}
	\vspace{-3mm}
	\label{fig_height_vs_width}
\end{figure}

{\bf The symbol width and channel capacity.} Our decoding method enables us to obtain binary data. But a designer willing to use this new channel would need more information to assess the feasibility of a potential application. For example, considering that the emitters (lights in buildings, streets or roads) have a fixed output power and height:
\begin{itemize}
	\item What symbol width should the designer use on objects to be able to decode information? 
	\item And given this symbol width, what channel capacity can the designer expect? 
\end{itemize}
Depending on the symbol width and the receiver's FoV, inter-symbol interference may appear. A wider symbol width makes the system more resilient to interference, see Fig.~\ref{fig_FOV}, but it reduces the amount of information that can be encoded on the surface of the object. A narrow symbol width has the opposite trade-off. To provide some insights we gradually change the receiver's and emitter's height from 20~cm to 55~cm. For each of these heights we test packets with different symbol widths, ranging from 1.5~cm to 7.5~cm. The {objects carrying the} packets are moved starting from a slow speed up to the maximum speed that allows the packets to be decodable. With this basic setup we identify two important trends:

\emph{Symbol width.} Fig.~\ref{fig_height_vs_width}(a) shows that there is a decodable region with a linear relationship between the maximum height of the emitter/receiver and the symbol width. Assuming a constant power at the light source, there will be a height beyond which the receiver will not be able to decode information, no matter how long the symbol width is. 

\emph{Throughput.} For a given emitter/receiver setup, the throughput is a function of the symbol width and the speed. Fixing one of these parameters determines the other. Using a constant speed of 8~cm/s, we have identified the narrowest symbol width that makes the packet decodable. Based on these experiments, we show that the channel's capacity decreases exponentially with the receiver's height in Fig.~\ref{fig_height_vs_width}(b). 

Trends like the ones exposed above are important for the design of applications. Getting receivers as close as possible to moving objects would enable 
(i) smaller objects to be monitored (cf. the linear improvement in Fig.~\ref{fig_height_vs_width}(a)), and (ii) exponentially more information (cf. Fig.~\ref{fig_height_vs_width}(b)).

\begin{figure}[!t]
	\begin{minipage}[t]{0.485\linewidth}
		\includegraphics[width=\columnwidth]{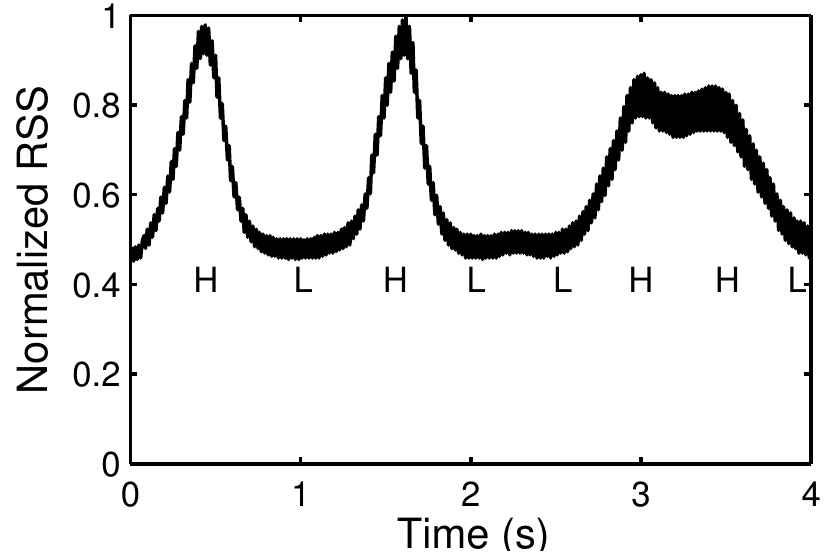}
		\vspace{-6mm}
		\caption{Signal received \newline under incandescent bulb.}
		\label{fig_incandescent}
	\end{minipage}%
	\hfill%
	\begin{minipage}[t]{.485\linewidth}
		\includegraphics[width=\columnwidth]{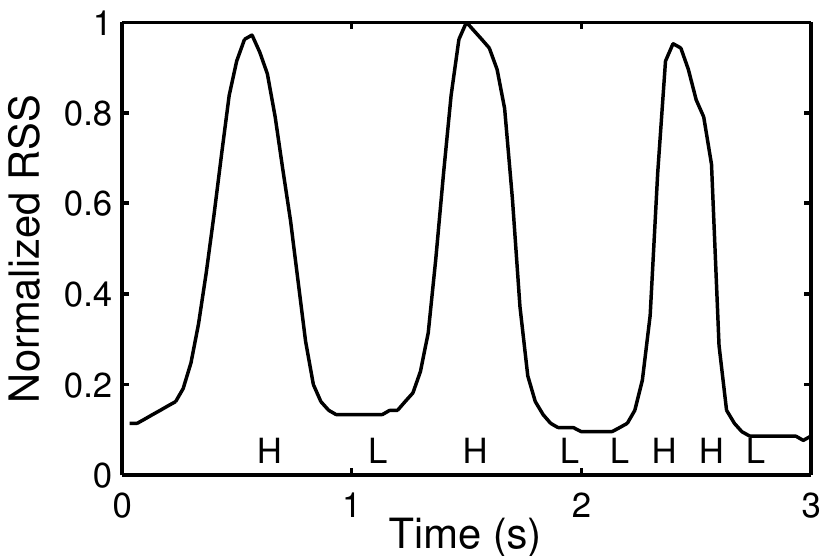}
		\vspace{-6mm}
		\caption{Signal received \newline under variable speed.}
		\label{fig_id_with_distortion}
	\end{minipage} 
	\vspace{-4mm}
\end{figure}

{\bf Impact of other light sources.} We also perform experiments using standard fluorescent lights in our ceiling. The results are shown in Fig.~\ref{fig_incandescent}. Here the lights' height is 2.3~m and the receiver's is 0.2~m.
The decoding method still works in these settings but there are some important points to highlight.
Note that because we have an illuminated area, the noise floor is higher, which leads to a smaller difference between the \texttt{HIGH} and \texttt{LOW} symbols compared to our dark-room experiments. There is also a larger variance in the signal, `thicker lines' than those in Fig.~\ref{fig_two_ids}, which is due to the AC power supply~\cite{Kuo2014VLCS}. The ceiling lights are significantly further away than the LED lamp used in our prior experiments, but due to their higher output power, we can still decode information. An extreme case highlighting the relationship between high output power and coverage is presented in Section~\ref{sec_app}, where we use the sun.

This experiment exposes an important trade-off in our channel. In traditional communication systems, the transmitter and the sources creating interference are normally independent, in our case they are the same. A light source with a high output power will increase the amount of light reflected to the receiver (\texttt{HIGH} symbol) but it also increases the noise floor.

\subsection{Channel Distortion: Variable Speed}
Thus far we have assumed an ideal scenario leading to a clean signal. In practice many events can distort the signal, such as objects moving at variable speed.
In these scenarios decoding the data may not be possible, but a plausible alternative is to transform the decoding problem into a classification problem.
We could compare the distorted signal against a database of clean signals (obtained under ideal scenarios, cf. Sec.~\ref{sec_ideal_scenarios}) to see which one is the best match. Clearly, in this case we will not be able to use $2^N$ codes. We will be constrained to use far less codes making sure that their inter-Hamming distances are maximized to have codes that are as different as possible from each other. 

We now showcase a scenario where channel distortion is caused by objects moving at variable speeds. While many signal processing techniques could be used for classification problems, we use Dynamic Time Warping (DTW) to showcase our basic idea. DTW is a method used in many areas to measure the similarity of two signals. We use the two clean signals in Fig.~\ref{fig_two_ids} as the baselines for comparison. For the distorted signal we use the same packet as the one in Fig.~\ref{fig_two_ids}(b), but we change its speed in the middle of the decoding process. This object moves at a certain speed when its first half (preamble) passes the receiver, and the speed is doubled when the second half (Data field) passes by. Figure~\ref{fig_id_with_distortion} depicts the RSS of this distorted signal. The decoding method presented in the previous subsection leads to an erroneous symbol sequence when used on the distorted signal, ``\texttt{HLHL.HL}", instead of the correct one ``\texttt{HLHL.LHHL}". With DTW, the normalized distances between the signals in Fig.~\ref{fig_id_with_distortion} and Fig.~\ref{fig_two_ids}(a), and between Fig.~\ref{fig_id_with_distortion} and Fig.~\ref{fig_two_ids}(b) are  326 and 172, respectively (as a reference, the distance between the signal in Fig.~\ref{fig_id_with_distortion} and itself is 131). Therefore, the distorted packet in Fig.~\ref{fig_id_with_distortion} is classified as belonging to the same group as the packet in Fig.~\ref{fig_two_ids}(b), which is correct.

\begin{figure}[!b]
	\centering
	\vspace{-2mm}
	\includegraphics[width=.75\columnwidth]{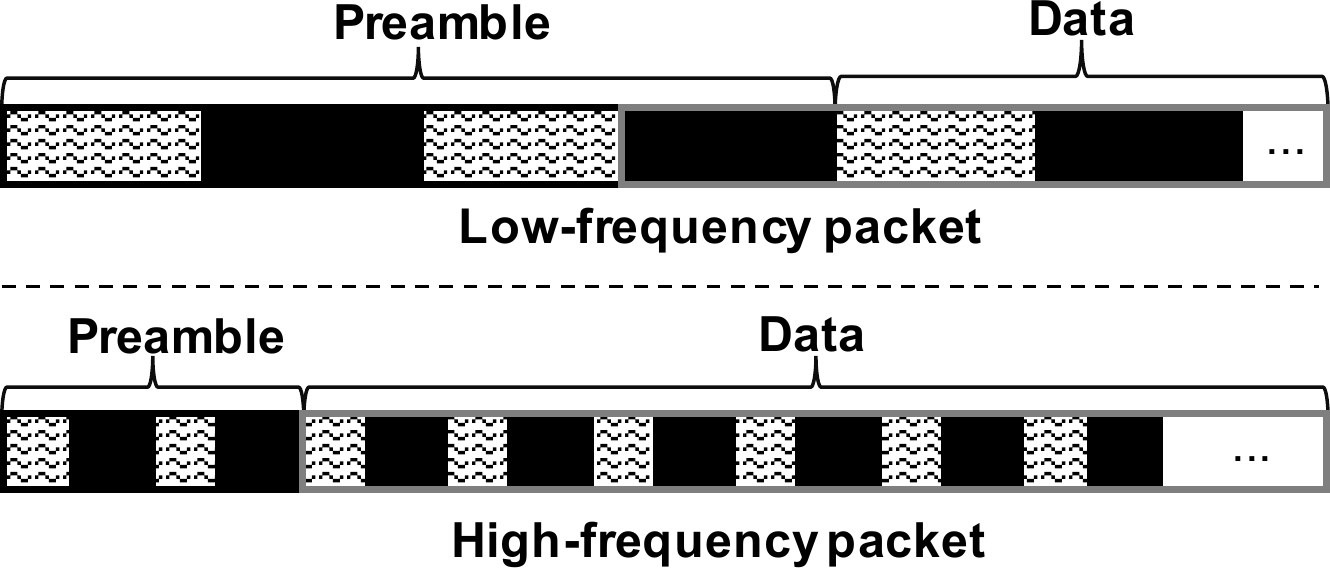}
	\vspace{-2mm}
	\caption {Illustration of the low/high-frequency packets.}
	\vspace{-1mm}
	\label{fig_hl_frq_pkt}
\end{figure}

\subsection{`Packet' Collisions: Frequency Domain}
In the previous two subsections, we assume there is only one packet moving under the FoV of the receiver. This approach works well for scenarios with `constrained' mobility, such as car lanes or tracks, but this assumption may not hold in less structured scenarios.
We now investigate the case where two packets pass through the FoV at the same time. We consider high- and low-frequency packets. A high-frequency packet is one with narrow symbol widths, so the received signal changes fast; and a low-frequency packet is one with wide symbol widths (slow-changing signal). 

\def \width{0.49} 
\begin{figure}[!t]
	\subfigure[{\tt Case1}- received signal] {\includegraphics[width=\width\columnwidth]{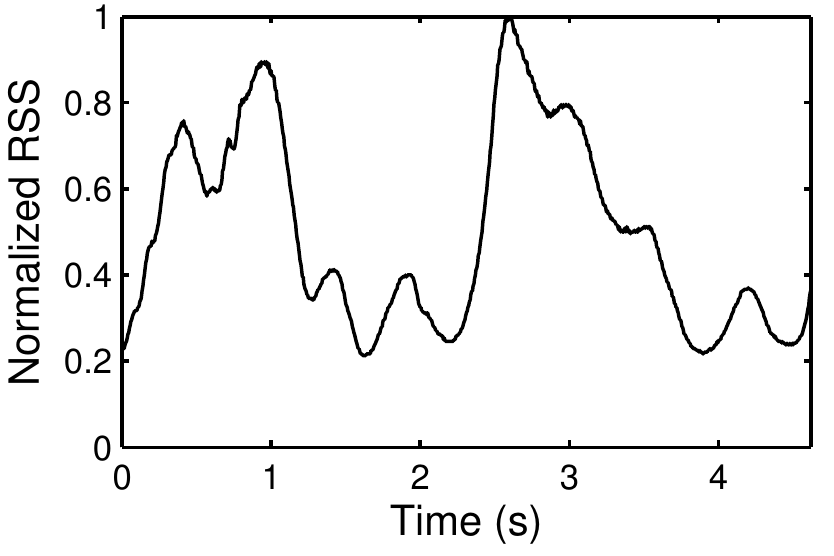}}
	\hfill
	\subfigure[{\tt Case1}- signal after FFT] {\includegraphics[width=\width\columnwidth]{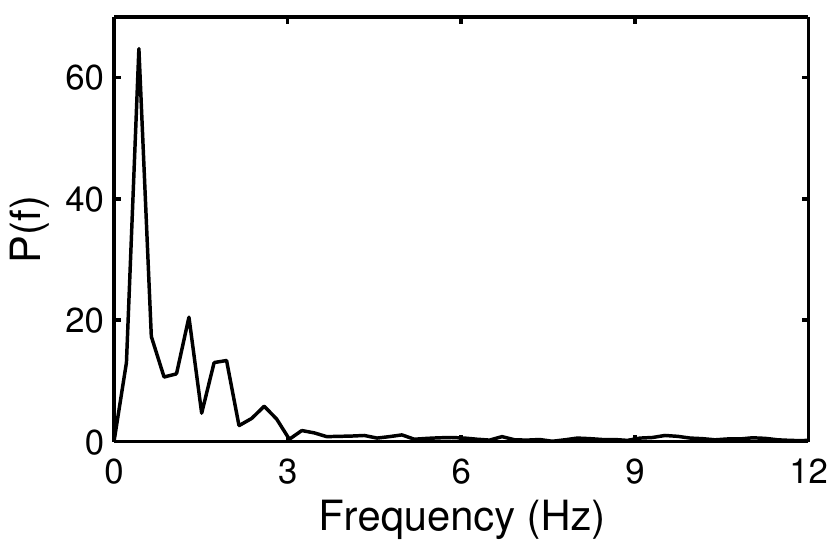}}
	
	\subfigure[{\tt Case2}- received signal]{\includegraphics[width=\width\columnwidth]{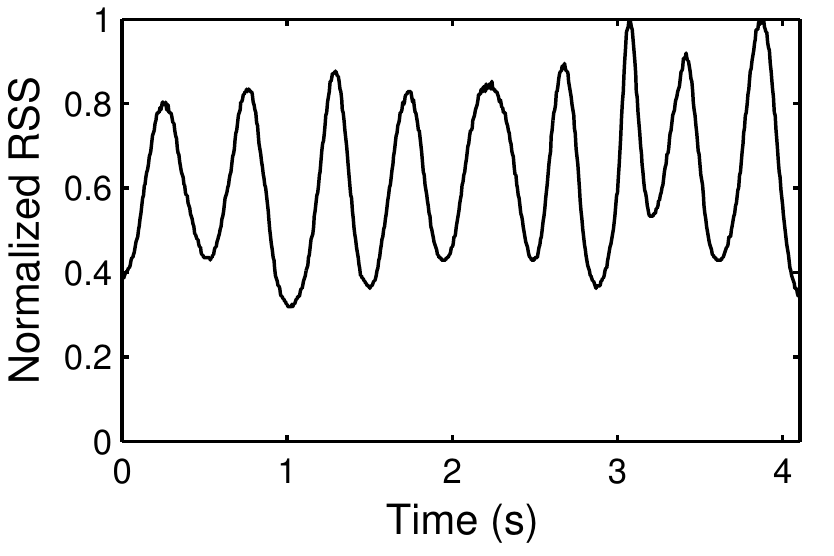}}
	\hfill
	\subfigure[{\tt Case2}- signal after FFT] {\includegraphics[width=\width\columnwidth]{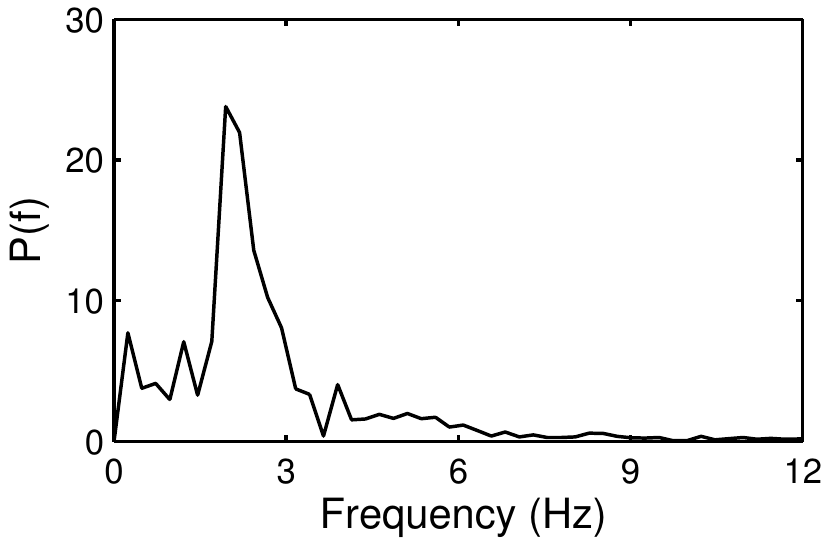}}
	
	\subfigure[{\tt Case3}- received signal] {\includegraphics[width=\width\columnwidth]{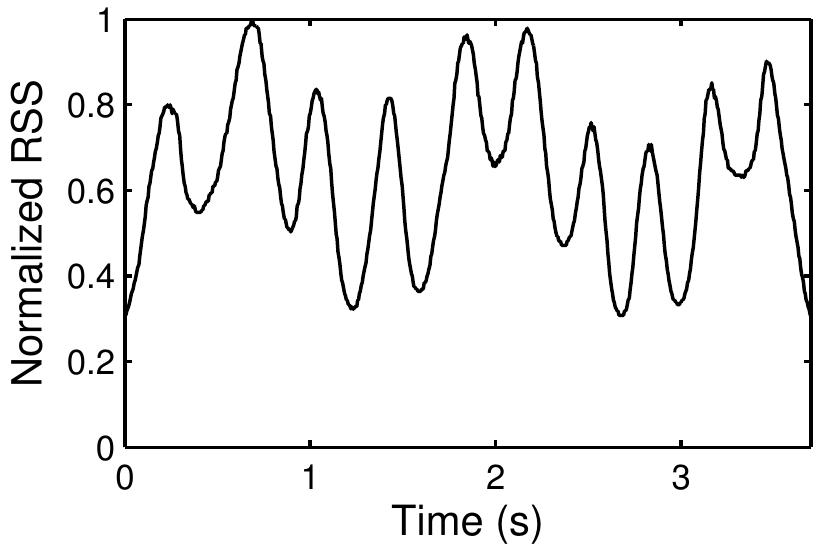}}
	\hfill
	\subfigure[{\tt Case3}- signal after FFT] {\includegraphics[width=\width\columnwidth]{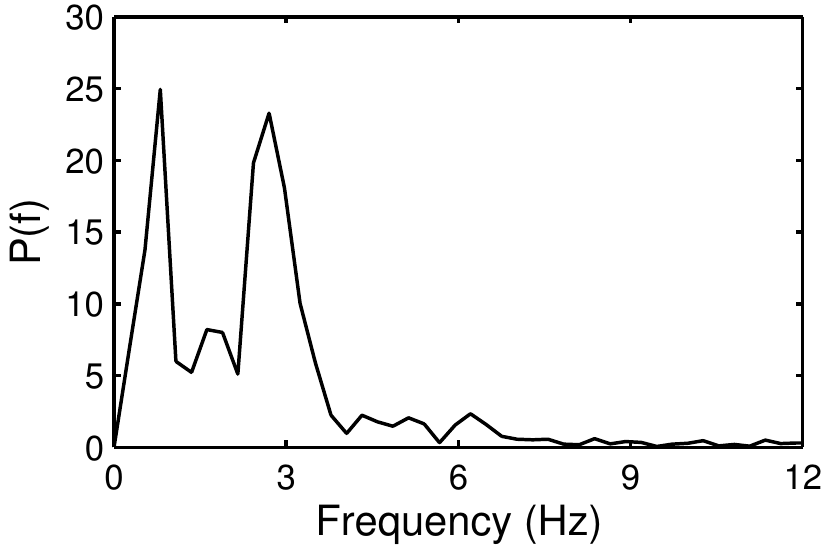}}
	\vspace{-2mm}
	\caption {Overlapping signals and their FFT.}
	\vspace{-4mm}
	\label{fig_multi_tag}
\end{figure}

In our tests both packets have the same length, their patterns are shown in Fig.~\ref{fig_hl_frq_pkt}. We carry out three different tests: 
\begin{itemize}
	\item {\tt Case1} --
	the low-frequency packet dominates the reflected light towards the FoV of the receiver; 
	\item {\tt Case2} -- the two packets exchange their positions (the high-frequency packet becomes the dominating signal);
	\item {\tt Case3} -- the two packets share equally the receiver's FoV (no dominant packet).
\end{itemize}

The RSS captured at the receiver for these three cases are shown in Fig.~\ref{fig_multi_tag}(a), (c) and (e). For {\tt Case1} and {\tt Case2}, we can decode bits using the method presented in Sec.~\ref{sec_ideal_scenarios}. For {\tt Case3} however, we cannot get accurate information from either method, decoding or DTW.

To obtain partial information for {\tt Case3}, we use Fast Fourier Transforms (FFT) to analyze collisions in the frequency domain. The frequency spectrums are shown in Fig.~\ref{fig_multi_tag}(b), (d) and (f). In Fig.~\ref{fig_multi_tag}(b) and (d), we can observe a single dominant frequency, which explains why it was easy to decode information in the time domain. But the undecodable signal in Fig.~\ref{fig_multi_tag}(e) also benefits from the FFT, because we can detect the presence of two different types of object.

These results however only scratch the surface of the problem. For a thorough investigation, setups such as having more than two packets under the FoV or having them pass with different speeds need to be addressed to quantify the interplay of the FoV and different collision scenarios.

\begin{figure}[!t]
	\begin{minipage}[t]{0.57\linewidth}
		\vspace{-31.5mm}
		\resizebox{\columnwidth}{!}{
			\begin{tabular}{|c|c|c|}
				\hline
				& {\bf Saturation} & {\bf Sensitivity}   \tabularnewline \hline 	
				PD ({\tt G1}) & 450~lux  & 1 \tabularnewline \hline
				PD ({\tt G2}) & 1200~lux  & 0.45 \tabularnewline \hline
				PD ({\tt G3}) & 5000~lux & 0.089  \tabularnewline \hline
				LED &  35000~lux & 0.013 \tabularnewline  \hline
		\end{tabular}}
		\caption{Supported noise floor of the optical receivers used in this work. The sensitivities are normalized to that of the PD with gain control {\tt G1}.}
		\label{fig_noise_floor} 
		\vspace{-4mm}
	\end{minipage}%
	\hfill%
	\begin{minipage}[t]{.4\linewidth}
		\centering
		\includegraphics[width=.9\columnwidth]{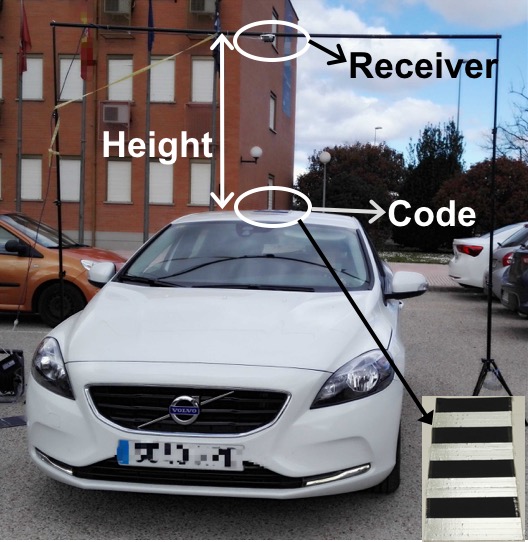}
		\vspace{-2mm}
		\caption{Setup of the outdoor application.}
		\label{fig_setup_outdoor_app}
		\vspace{-4mm}
	\end{minipage} 
\end{figure}

\subsection{Noise Floor: tradeoff between PD and LED as a Receiver}\label{subsec:noise_floor}

Until now our focus has been on obtaining information  at different levels: decoding (clean channel) and waveform mapping (distorted channel). But these evaluations do not consider explicitly the drastic changes in the noise floor, which cause photodiodes (PDs) to get easily saturated under strong light conditions, especially outdoor~\cite{Liu:2011}. Given that we exploit ambient light as the `emitter'  we need a receiver that can work in a wide range of ambient illuminance conditions. 

To obtain a solid receiver, we propose to use conjunctly a PD and an LED acting as a receiver (RX-LED), both shown in Fig.~\ref{fig:openvlc}. LEDs acting as receivers have been used before in VLC systems~\cite{Dietz,giustiniano_wd,Schmid2013,Wang2015HotWireless} and have different optical properties compared to PDs. In particular, we want to exploit two key properties of RX-LEDs: narrow FoV and narrow optical bandwidth. These two properties lead to lower sensitivities (which is usually not good), but they also increase the resilience to saturation. To quantify this tradeoff for the specifics of our channel we perform the following experiments. We use three different gain controls for the PD, from a high value {\tt G1} (high sensitivity and long range but easily saturated) to a low value {\tt G3} (opposite effect). To increase the sensitivity of the RX-LED we decide to operate it in photovoltaic mode (as solar cells) that minimizes the effect of dark current (current generated in the absence of ambient light). The results are shown in Fig.~\ref{fig_noise_floor}. We have two observations:

\textit{Saturation}: the PD at gain control level {\tt G1} saturates at 450 lux, which maps roughly to a medium illuminated room. At {\tt G3}, the PD works for noise floors up to 5000 lux. But outdoor scenarios during the day can easily go above 10 klux. The RX-LED, instead, can work when the noise floor is up to 35,000 lux and is thus more suitable for outdoor scenarios.

\textit{Sensitivity}: the RX-LED is less sensitive than the PD, which will affect their ability to decode data at low RSSs, such as indoor environments with low ambient illuminance. 

Concluding, a receiver with two optical components (PD and RX-LED) can alleviate the noise floor problem by properly selecting the component that provides reliable passive communication for the given ambient light conditions.

\section{Application Evaluation} \label{sec_app}

\begin{figure}[t!]
	\begin{minipage}[b]{.485\columnwidth}
		\centering
		\subfigure
		{\includegraphics[width=.95\columnwidth,height=.41\columnwidth]{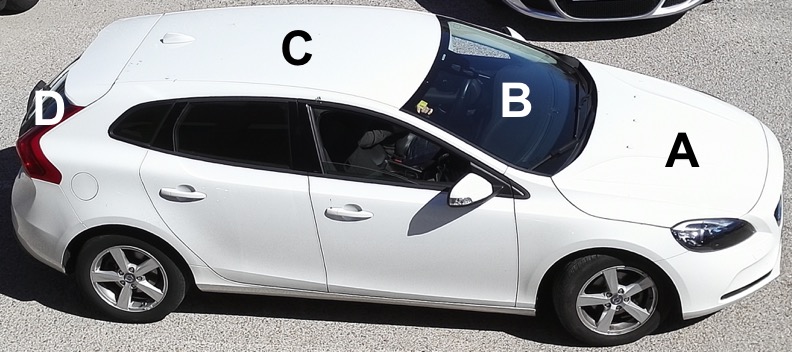}}
		\subfigure {\includegraphics[width=\columnwidth]{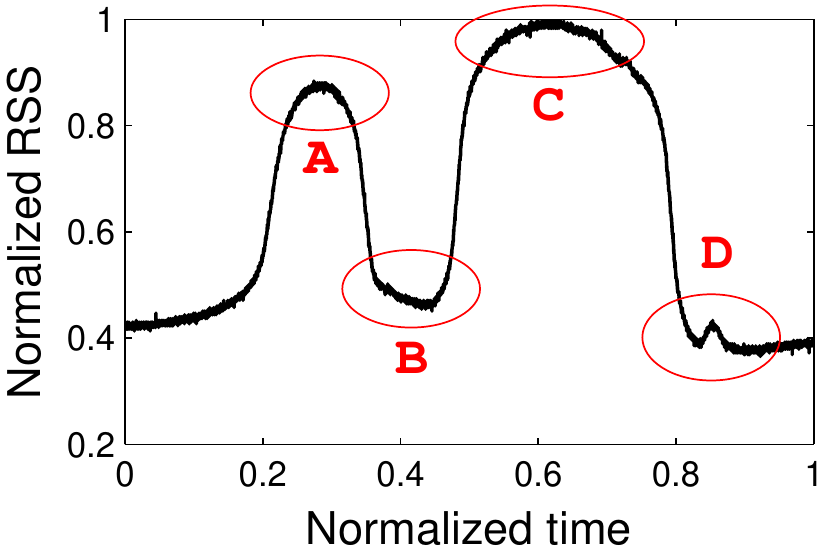}}
		\vspace{-3mm}
		\caption{Top: Volvo V40; \newline bottom: captured signal.}
		\vspace{-2mm}
		\label{fig_car_volvo_v40}
	\end{minipage} 
	\hfill
	\begin{minipage}[b]{.485\columnwidth}
		\centering
		\subfigure {\includegraphics[width=.95\columnwidth,height=.41\columnwidth]{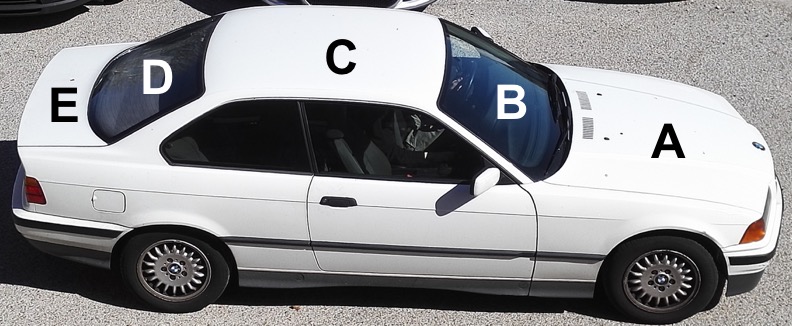}}
		\subfigure {\includegraphics[width=\columnwidth]{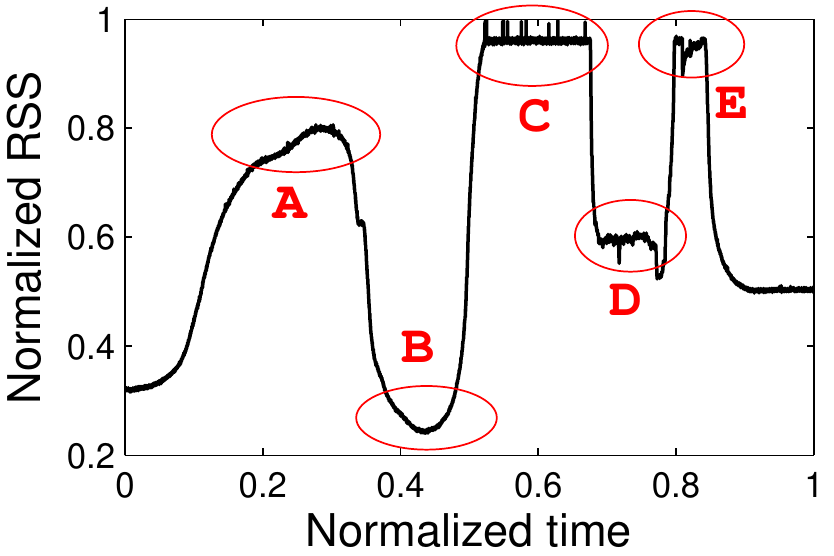}}
		\vspace{-3mm}
		\caption{Top: BMW3; \newline bottom: captured signal. }
		\vspace{-2mm}
		\label{fig_car_bmw_3}
	\end{minipage} 
\end{figure}

\begin{figure}[t]
	\subfigure[Noise floor: 450~lux] {\includegraphics[width=0.49\columnwidth]{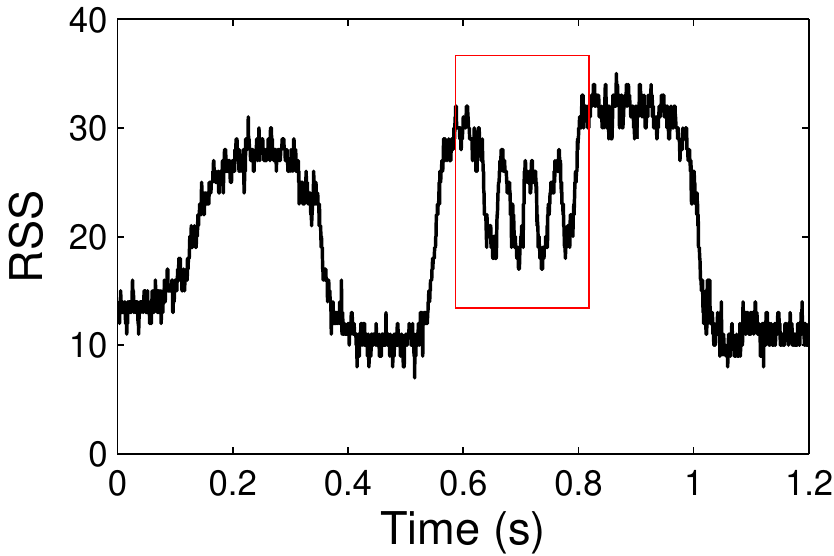}}
	\hfill
	\subfigure[Noise floor: 100~lux] {\includegraphics[width=0.49\columnwidth]{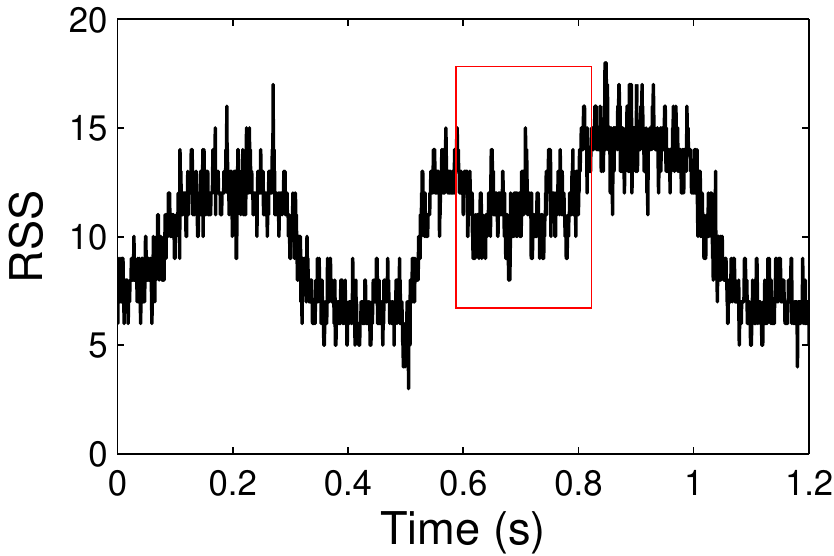}}
	\vspace{-2mm}
	\caption {LED as RX. Car's speed is 18~km/h, the receiver's height is 25~cm, and the code is `HLHL.HLHL'.}
	\vspace{-2mm}
	\label{fig_d25cmLED}
\end{figure}

\begin{figure}[t!]
	\subfigure[Noise: 100~lux; w/o shield.] {\includegraphics[width=0.49\columnwidth]{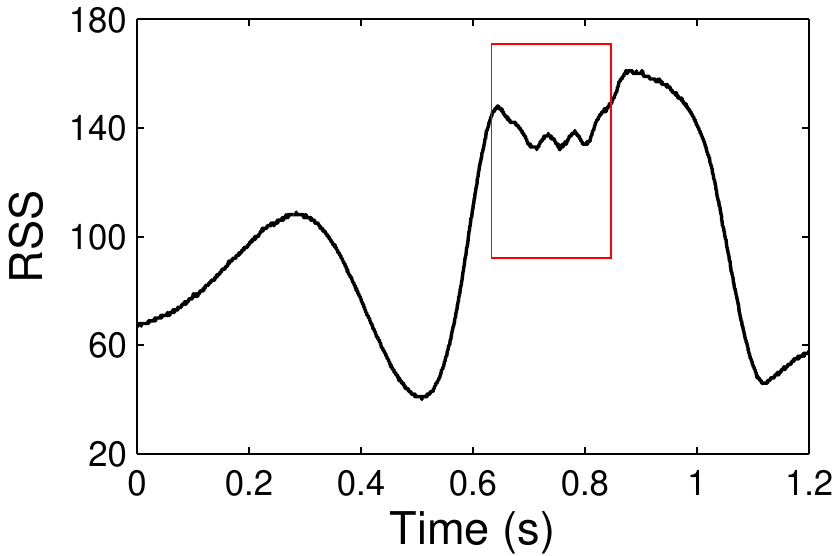}}
	\hfill
	\subfigure[Noise: 100~lux; w/ shield.] {\includegraphics[width=0.49\columnwidth]{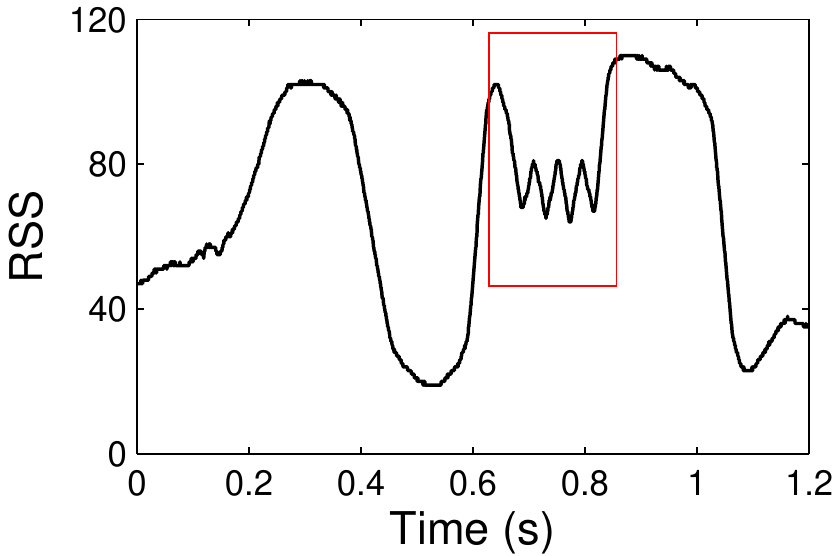}}
	\vspace{-2mm}
	\caption {PD as RX. Car's speed is 18~km/h, the receiver's height is 25~cm, and the code is `HLHL.HLHL'.}
	\vspace{-2mm}
	\label{fig_d25cmPD}
\end{figure}

\begin{figure*}[t]
	\subfigure[Height: 75 cm; noise floor: \newline 6200 lux; code: `HLHL.HLHL'.]
	{\includegraphics[width=0.24\textwidth]{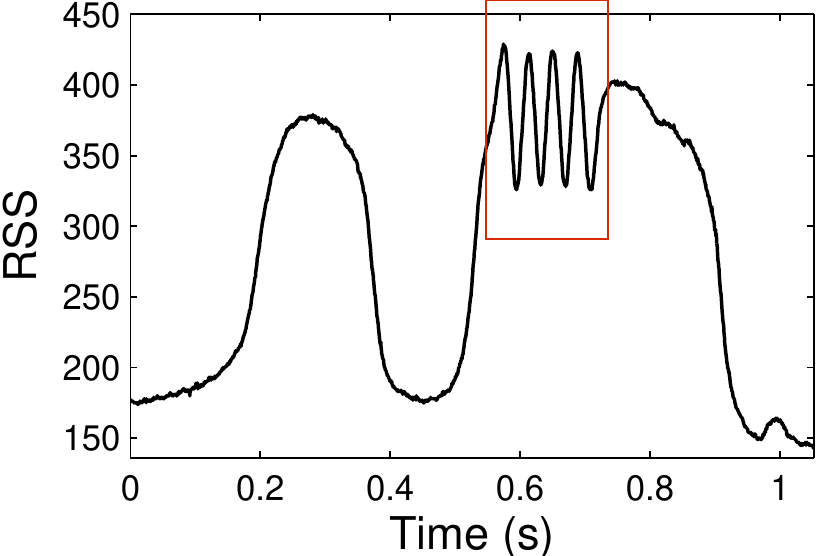}}
	\hfill
	\subfigure[Height: 100 cm; \ noise floor: \newline 3700 lux; \   code: `HLHL.HLHL'.]
	{\includegraphics[width=0.24\textwidth]{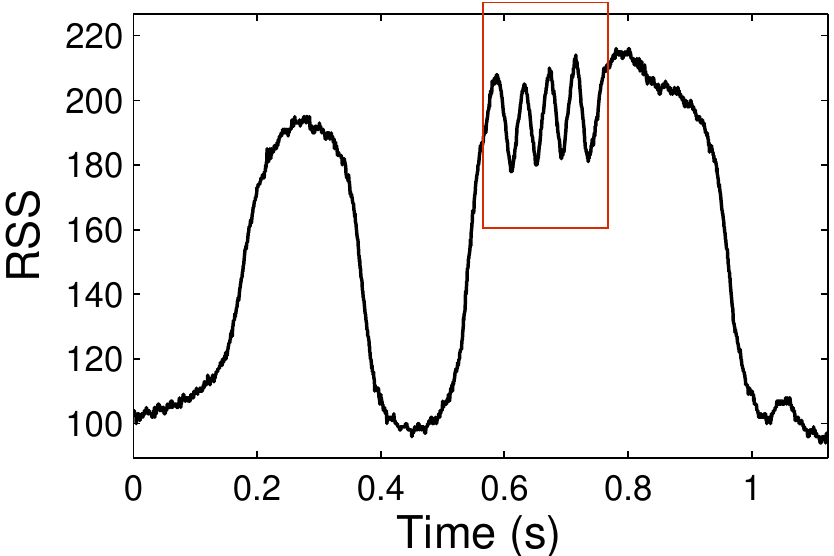}}
	\hfill
	\subfigure[Height: 100 cm; noise floor: 5500 lux; \ code: `HLHL.LHHL'.]
	{\includegraphics[width=0.48\textwidth]{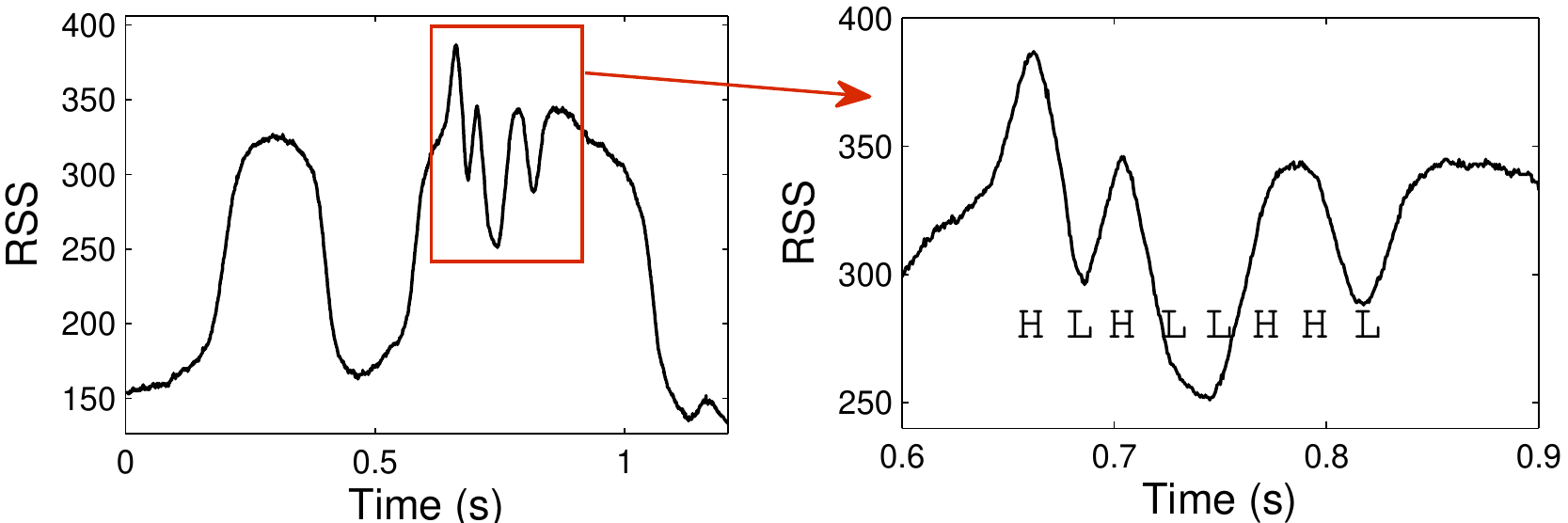}}
	\vspace{-2mm}
	\caption {Car's speed is 18~km/h. Two different types of code and different distance between the car and the receiver.}
	\vspace{-2mm}
	\label{fig_speed100cmLEDID1}
\end{figure*}

We now evaluate our system with an outdoor application. We place a `packet' on the roof of a car and attach the receiver to a pole supporting structure, cf. Fig.~\ref{fig_setup_outdoor_app}. The receiver uses two optical components: a PD and a RX-LED. The goal is to decode the packet information as the car passes by. The symbol width is 10\,cm and the sampling frequency of the receiver is set to 2K samples/sec. The experiments are carried out in cloudy days at noon and late afternoon {to study different sun positions and intensities}.

\subsection{Baseline: Car's Shape Detection}

We use two cars in our experiments: a Volvo V40 and a BMW series 3. As baseline experiments, we drive the cars under the receiver without any packet on the roof. The top part of the cars have two different materials, metal and glass, with different lengths and shapes. Thus, their optical signatures should be unique. Figs.~\ref{fig_car_volvo_v40} and~\ref{fig_car_bmw_3} show the baseline signals of the cars captured with the RX-LED. We can observe that the metal parts of the cars --hoods (A), roofs (C) and trunks (E)-- reflect much more light (peaks) than the front and rear windshields (B and D). The different designs of the cars are also accurately reflected by their waveforms.

The ability to detect the shape of the car with the RX-LED allows us to use the car's optical signature as a {\it long-duration-preamble} of the packet, indicating when the receiver needs to get ready to decode information. 
Due to space limitations, {but without the loss of generality}, we now only report the experiments with the Volvo V40 at a  speed of 18\,km/h. 

\vspace{-1mm}
\subsection{Mild Illuminated Environment}
We perform tests for various heights of the receiver and car speeds. To decode the information, we use the same method described in Section~\ref{sec_evalu} but in two phases:
\begin{itemize}
	\item We first look for the long-duration-preamble based on the car's shape (by detecting the hood `peak' and windshield `valley', cf. Fig.~\ref{fig_car_volvo_v40});
	\item We then perform the decoding algorithm in Sec.~\ref{sec_ideal_scenarios}.
\end{itemize}

\textbf{Impact of noise floor.}
A mild illumination requires a lower height to ensure the system works well. In Fig.~\ref{fig_d25cmLED}, we observe that when the noise floor is about 450~lux, our prototype works well with the RX-LED up to a height of 25~cm. When the noise floor decreases to 100~lux, we cannot decode the information anymore, cf. Fig.~\ref{fig_d25cmLED}(b). The reason behind this phenomenon is that our system harnesses the ambient light to modulate its information. If the ambient light is too weak, the modulated information can not travel too far due to the light's attenuation (cf. Section~\ref{subsec:noise_floor}).

Since the PD is more sensitive than RX-LED (cf. Fig.~\ref{fig_noise_floor}), we use it with gain G2 to overcome the low illuminance conditions. Fig.~\ref{fig_d25cmPD}(a) shows that while the signal is cleaner than the one obtained with RX-LED, the information is still not decodable. The reason is that the PD has a large FoV, thus the car's metal roof adds interference at the receiver. By reducing the PD's FoV with a small physical cap (1.2$\times$1.2$\times$2.8\,cm), we filter out much of the interference and decode the information, as shown in Fig.~\ref{fig_d25cmPD}(b), regardless of the RSS drop resulting from the smaller impinging light on the receiver.

\vspace{-1mm}
\subsection{Well Illuminated Environment}

Given that the PD can easily saturate with higher noise floor (cf. Section~\ref{subsec:noise_floor}), we evaluate the RX-LED for passive communication under well illuminated conditions.

\textbf{Supported car's speed.} We fix the height of the receiver to 75~cm and drive the car at 18~km/h. The noise floor is around 6200~lux during the experiments. The results are shown in Fig.~\ref{fig_speed100cmLEDID1}(a). We can observe that the captured signal is very clear, and thus, the information is easy to decode. The achieved throughput is around 50 symbols/s.

\textbf{Maximal height.} We then increase the height of the receiver to 100~cm. The noise floor is about 3700~lux. Under these circumstances, we can decode the information easily.  Note that the RSS in Fig.~\ref{fig_speed100cmLEDID1}(b)  is smaller than that in Fig.~\ref{fig_speed100cmLEDID1}(a). This is because in Fig.~\ref{fig_speed100cmLEDID1}(b), the height is higher and the ambient light is weaker compared to Fig.~\ref{fig_speed100cmLEDID1}(a). Finally, the result with the car carrying a different packet code and under a different noise floor is shown in Fig.~\ref{fig_speed100cmLEDID1}(c). The packet is successfully decoded in this case as well.

\section{Discussion and Conclusion} \label{sec_discu}

We have presented a passive communication system that harnesses ambient light to convey information. 
The three block elements of our system, the emitter, the `packet' and the receiver, are sustainable in its conception. We identified fundamental challenges and evaluated a prototype in an outdoor parking lot. While the obtained results are encouraging, more research is necessary. 
(1) \emph{Encoding dynamic data.} Our system can now encode static data in each packet, targeting at low footprint. Encoding dynamic information is feasible by adopting advance materials whose reflection is adjustable (e.g. E-ink screens or LCD shutters).
(3) \emph{Maximal supported speed of an object.} This is mainly determined by the PD's response time to light changes and the receiver's sampling rate. We will exploit this in a follow-up work.
(4) \emph{Reflected light intensity.} Our system exploits reflected light to convey information, whose intensity is normally weaker than direct light. To improve the performance, we expect to benefit from next-generation optical receivers in VLC or advanced diffraction gratings (e.g., holographic gratings~\cite{Yetisen2013}). (5) \emph{Networking.} If the receivers in our system are networked, then they can share the information about the tracked objects and thus could improve the system's performance. How to connect these low-end receivers efficiently is a challenge in our system. These challenges are left for future research, yet this work provides a first step for new types of  passive communication with visible light.

\section*{Acknowledgement}
This research work was funded in part by Ministerio de Econom\'ia y Competitividad grant TEC2014-55713-R and by the Madrid Regional Government through the TIGRE5-CM program \newline (S2013/ICE-2919). Our sincere thanks to the reviewers and our shepherd Monia Ghobadi for all of their great comments and constructive feedbacks.

\balance
\bibliographystyle{acm}
\bibliography{bibliography_vlc}

\end{document}